\documentclass[aps,twocolumn,prd,amssymb,nofootinbib,floatfix,superscriptaddress]{revtex4}
\usepackage{epsfig}
\usepackage{amssymb}
\usepackage{amsmath}
\usepackage{amsfonts}

\usepackage{bm}
\usepackage{revsymb}
\usepackage{graphicx,epsfig}
\usepackage{placeins}
\newcommand{\be}{\begin{equation}}
\newcommand{\ee}{\end{equation}}
\newcommand{\bea}{\begin{eqnarray}}
\newcommand{\eea}{\end{eqnarray}}
\newcommand{\bes}{\begin{subequations}}
\newcommand{\ees}{\end{subequations}}

\begin{document}
\interfootnotelinepenalty=10000

\title{Junction conditions at spacetime singularities}

\author{Eran Rosenthal}
\email{eranr@astro.cornell.edu}
\affiliation{Center for Radiophysics and Space Research, Cornell  University, Ithaca, New York, 14853}


\begin{abstract}
A classical model for the extension of singular spacetime geometries  across their singularities is presented.
The regularization introduced by this model is based on the following observation. 
Among the geometries that satisfy Einstein's field equations there is a class of geometries, with certain singularities, where the 
components of the metric density and their partial derivatives 
remain finite in the limit where the singularity is approached.  
Here we exploit this regular behavior of the metric density and 
elevate its status to that of a fundamental variable -- from which the metric is constructed. 
We express Einstein's field equations as a set of equations for the metric density, and postulate junction 
conditions that the metric density satisfies at singularities. 
Using this model we extend certain geometries across their singularities.
The following examples are discussed: radiation dominated Friedmann-Robertson-Walker Universe, 
Schwarzschild black hole, Reissner-Nordstr\"{o}m  black hole, and certain Kasner solutions. 
For all of the above mentioned examples 
we obtain  a unique extension of the geometry beyond the singularity.
\end{abstract}

\maketitle

\section{Introduction}

General relativity (GR) is in excellent agreement with all
experimental and observational tests of gravity. 
However, when a GR solution develops a singularity it is sometimes impossible to use Einstein's field equations  
to extend the 
geometry beyond the hypersurface where the singularity resides. 
The standard lore is that 
this incompleteness should be  resolved by a quantum gravity 
theory that may become important when the curvature length scale 
is of the order of Planck length. 
It is possible that quantum-gravity phenomena can smooth out singularities and 
thereby resolve the nonextendibility of singular GR solutions.
Alternatively, it is possible that 
the correct description of gravity in the vicinity of singularities 
requires first an extension or a modification 
of classical GR (e.g. classical string theory introduces 
modifications to GR \cite{GSW}). In this 
case, quantization of gravity can take place only after 
 our classical understanding of gravity has been modified \cite{HM}. 
Therefore, there is a good motivation to extend or modify GR 
 so that it would not breakdown at singularities.  Moreover, 
quantum theories can inherit singularities   
that appear in their corresponding classical theories 
(e.g. a charged point particle gives rise to an infinite electrostatic energy  in classical electrodynamics, a related ultraviolet divergent behavior also appears in quantum electrodynamics). 
Therefore, handling spacetime singularities at the classical 
level may provide a useful preliminary step towards a more regular quantum gravity analysis  of singularities.   
 Furthermore, this classical analysis could shed light on the extension of spacetime geometries beyond their singularities.

In this manuscript we introduce a classical model for gravity    
that retains its predictive power for a class of spacetime singularities. 
In the construction of the model we make sure that the following 
requirements are satisfied:  
First, to ensure agreement with observational constraints, 
we demand that except for singularities 
the model would coincide with GR. 
Second, we demand that the model 
would be able to provide predictions  
for the extension of certain spacetime geometries beyond their singularities.

Below we use our model to extend certain GR solutions 
across their singularities. 
The following examples 
are studied:  Friedmann-Robertson-Walker Universe (FRW) , 
Schwarzschild black hole, charged Reissner-Nordstr\"{o}m  black hole (RN), and Kasner Universe. 
For other classical approaches to singularities 
see Penrose \cite{penrose}  and Tod \cite{tod}.

The regularization that we introduce is based on the following observation. 
There is a class of spacetime singularities where 
the components of the metric density (in certain coordinates)  
attain  finite values in the limit where the singularity is approached. 
As an example let us consider a 
 Schwarzschild black hole. In the Schwarzschild coordinates the metric reads 
$g^{Sch}_{\mu\nu}={\rm diag}(-b/r,r/b,r^2,r^2\sin^2\theta)$, where $b=r-2M$, and 
 M denotes the mass of the black hole. 
In the limit $r\rightarrow 0$, the Kretschmann scalar  
$R^{\alpha\beta\gamma\delta}R_{\alpha\beta\gamma\delta}$ 
diverges  thereby indicating that there is a spacetime singularity at $r=0$.     
Here we focus attention on  the contravariant metric density $\frak{g}_{Sch}^{\alpha\beta}$ which is equal to 
 the product  $[{-{\rm det}(g_{\mu\nu}^{Sch})}]^{1/2}g_{Sch}^{\alpha\beta}$.
In the  Schwarzschild coordinates,  it reads 
$\frak{g}_{Sch}^{\alpha\beta}={\rm diag}
[-b^{-1}r^3\sin\theta,rb\sin\theta,\sin\theta,\sin\theta^{-1}]$. 
Notice that in the limit of interest 
the components of $\frak{g}_{Sch}^{\alpha\beta}$ and their partial derivatives  remain finite, 
though the complete matrix $\frak{g}_{Sch}^{\alpha\beta}$ becomes degenerate. 
In this paper we exploit this non-divergent behavior of the metric density 
and elevate its status to that of a fundamental variable, 
while the metric becomes a constructed variable, which is constructed from the metric density. 
We express Einstein's field equations as a set of equations for the metric density 
and postulate junction conditions 
at the singularities. 
We use this framework to extend certain geometries across their singularities.
At the singularities we find that 
the components of the metric density are continuous 
and differentiable even though the geometry 
 remains ill-defined. 
In fact our model does {\em not} 
remove any physical singularity.  On the contrary, 
as in GR we find that the  Schwarzschild geometry becomes singular as $r\rightarrow 0$. 
However, in our framework the spacetime geometry becomes 
a constructed entity, which is constructed from a non-divergent fundamental quantity 
-- the metric density. 
Since the metric  is no longer a fundamental dynamical variable,  
 its singular behavior at a spacetime singularity 
does not obstruct the continuation 
of the solution beyond the singularity.

This paper is organized as follows: in Sec. \ref{model} the classical model is presented, in Sec. {\ref{examples}} 
we discuss few examples in detail, and  Sec. \ref{conclusions} provides conclusions.

\section {The Classical Model}\label{model}

Let us begin by 
defining the metric density $\frak{g}^{\mu\nu}$ to be a four-dimensional symmetric 
 contravariant tensor density \cite{density} of weight -1. 
In the standard GR formulation the metric is represented by a non-degenerate matrix.
Recall, however, that the metric density may become degenerate 
in the limit where a spacetime singularity is approached. For example, 
  the Schwarzschild metric density $\frak{g}_{Sch}^{\alpha\beta}$  becomes a degenerate matrix in the limit 
 $r\rightarrow 0$.  
Exploiting the non divergent behavior of $\frak{g}_{Sch}^{\alpha\beta}$, our first step is to extend the solution 
$\frak{g}_{Sch}^{\alpha\beta}$ in a continuous manner, and allow the metric density to attain 
its degenerate configuration in the limit. 
More broadly, a similar non-diverging behavior of the metric density 
also appears in other GR solutions; for example, 
in appropriate coordinates, the components of the metric densities of RN, Kerr, Kerr-Newmann, FRW, and 
Kasner solution, all 
remain finite at their corresponding singularities (see discussion below).
This means that within the space of GR solutions there is a class of singular solutions
that approach finite but degenerate configurations at their singularities.  
Our strategy is to include these degenerate metric density configurations in our formulation, and 
thereby construct a {\em continuous} extension of 
the space of GR solutions.

Notice that it is possible to introduce a coordinate transformation 
such that in the new coordinates the metric density $\frak{g}_{Sch}^{\alpha\beta}$ would diverge at the singularity. 
This divergent behavior 
is merely a coordinate singularity. In analogy with GR, we shall restrict attention to a class of 
nonsingular coordinates, where  the metric density and  its derivatives 
(see more details below) do not diverge.

From the metric density we construct the contravariant metric through
\be \label{metricdef}
g^{\alpha\beta}\equiv (-\frak{g})^{-1/2}\frak{g}^{\alpha\beta}\,.
\ee
Here
\be\label{det}
\frak{g}\equiv \det (\frak{g}^{\alpha\beta})\, .
\ee
Except for being the quantity from which the metric is derived, the metric 
density also has its own independent physical meaning.
In analogy with the metric, one can employ the metric density 
to construct scalars. These scalars are normally associated with a four-dimensional domain $D$.  
 For example, 
 the four volume of a domain $D$ is defined by
\be\label{v}
V\equiv\int_D \sqrt{-\frak{g}} d^4\,x\,.
\ee
The inner products between covector fields, and the inner product between 
vector fields are defined to be 
\begin{eqnarray}\label{dd}
&&(a_\alpha,b_\beta)\equiv\int_D a_{\alpha} b_{\beta} \frak{g}^{\alpha\beta} d^4x\,,\\\label{uu}
&&(u^\alpha,v^\beta)\equiv\int_D u^{\alpha} v^{\beta} \frak{g}_{\alpha\beta} d^4x\,.
\end{eqnarray}
Here the covariant metric 
density $\frak{g}_{\alpha\beta}$ is defined to be the adjoint matrix (also called the adjugate matrix)
of $\frak{g}^{\alpha\beta}$ 
\be\label{adjoint}
\frak{g}_{\alpha\beta}\equiv{\rm adj}(\frak{g}^{\alpha\beta})\,,
\ee
This means that $\frak{g}_{\alpha\beta}$  is equal to the transpose of the cofactor matrix of $\frak{g}^{\alpha\beta}$.
From definition (\ref{adjoint}) it follows that $\frak{g}_{\alpha\beta}$  
transforms as a covariant tensor density of weight $-1$ under a general coordinate transformation \cite{adjoint}, and furthermore
Eq. (\ref{adjoint}) guarantees that $\frak{g}_{\alpha\beta}$  remains well defined 
even if $\frak{g}^{\alpha\beta}$ becomes degenerate.
For example, in the Schwarzschild singularity both $\frak{g}_{\alpha\beta}^{Sch}$ and $\frak{g}^{\alpha\beta}_{Sch}$ as well as their partial derivatives (up to all orders), all have a well defined 
limit\footnote{We ignore coordinate singularities, e.g. at $\theta=0$, since these singularities
can be removed by a coordinate transformation.} 
as $r\rightarrow 0$. 
By contrast, the singular behavior of the Kretschmann scalar at the Schwarzschild singularity 
implies  that the metric must exhibit some kind of singular behavior at this limit\footnote {More precisely, 
since the diverging Kretschmann scalar is equal to a sum of products of terms that depend on the metric (including 
 the metric, its inverse, and their derivatives up to the second) it follows
that in all possible coordinates at least one of these metric dependent terms must  
diverge at the singularity.}.

 Notice that "observables" (\ref{v},\ref{dd},\ref{uu}) that are  constructed directly 
from the metric-density remain well defined 
even if domain $D$ contains  a singularity where $\frak{g}^{\mu\nu}$ takes the form of a nondiverging degenerate matrix. 
By contrast, a curve in $D$ that passes through such singularity may have an ill defined length.
From Eq. (\ref{adjoint}) we find that 
\be\label{relation1}
\frak{g}^{\alpha\mu} \frak{g}_{\mu\beta} =\frak{g}\,\delta^{\alpha}_{\beta}\,.
\ee

In analogy with the metric, one can use the metric density 
to construct evolution equations for other fields. Equations of motion that depend directly on 
 the metric density may retain their predictive power at singularities where the metric density becomes degenerate.
For example, in a fixed RN background the scalar wave equation  $(\frak{g}^{\alpha\beta}\phi_{,\alpha})_{,\beta}=0$ determines  
 the transmission of scalar waves through the RN singularity \cite{GKOS}.

Next we  define the covariant metric to be 
\be\label{ddmetricdef}
g_{\alpha\beta}\equiv -(-\frak{g})^{-1/2}\frak{g}_{\alpha\beta}\,.
\ee
It follows from Eqs. (\ref{metricdef},\ref{det},\ref{relation1},\ref{ddmetricdef}) that
for  $\frak{g}\ne 0$ we have 
\begin{eqnarray}\label{relations2}
&&\frak{g}=g\ , \ g^{\alpha\mu}g_{\mu\beta}=\delta^\alpha_\beta\,,\\\label{invrelation}
&&\frak{g}_{\alpha\beta}=-\sqrt{-g}g_{\alpha\beta}\ ,\ \frak{g}^{\alpha\beta}=\sqrt{-g}g^{\alpha\beta}\,.
\end{eqnarray}
Here $g\equiv\det(g_{\alpha\beta})$.

We now construct the equations of motion for the metric density. For this purpose  
we cast Einstein's field equations in a densitized
format. 
Using the standard Landau-Lifshitz formulation \cite{LL} and units where $G=c=1$ we obtain
\be\label{efe}
  (\frak{g}^{\mu\nu}\frak{g}^{\alpha\beta}- \frak{g}^{\mu\alpha} \frak{g}^{\nu\beta})_{,\alpha\beta}=
{16\pi}   \frak{T}^{\mu\nu}_{total} \,.
\ee
Here  $\frak{T}^{\mu\nu}_{total}=\frak{T}^{\mu\nu} +  \frak{t}_{LL}^{\mu\nu} $,
 $\frak{T^{\mu\nu}}=(-\frak{g})T^{\mu\nu}$, 
where $T^{\mu\nu}$  denotes the energy-momentum tensor depending on the matter 
fields and the metric density via Eqs. (\ref{metricdef},\ref{ddmetricdef}), 
 and
\begin{eqnarray}\nonumber
&&\frak{t}^{\mu\nu}_{LL}=\frac{1}{16\pi}
\left[ 2\frak{g}^{\mu[\nu}_{\ \ \ ,\lambda} \frak{g}^{\lambda]\omega}_{\ \ \ ,\omega}
+\frac{1}{2}\frak{g}^{-1}\frak{g}^{\mu\nu} \frak{g}_{\lambda\alpha} \frak{g}^{\lambda\beta}_{\ \ ,\rho} \frak{g}^{\rho\alpha}_{\ \ ,\beta}
\right. \\\nonumber
&&\left. 
-2\frak{g}^{-1}\, \frak{g}_{\ \ }^{\lambda(\mu} \frak{g}^{\ \ }_{\alpha\beta}  \frak{g}^{\nu)\beta}_{\ \ \ ,\rho}\frak{g}^{\alpha\rho}_{\ \ ,\lambda}
+\frak{g}^{-1}\frak{g}_{\lambda\alpha}\frak{g}^{\beta\rho}\frak{g}^{\mu\lambda}_{\ \ ,\beta}\frak{g}^{\mu\alpha}_{\ \ ,\rho}\right.\\\nonumber
&&\left.+\frac{1}{8}\frak{g}^{-2}
(2\frak{g}^{\mu\lambda}\frak{g}^{\nu\alpha}-\frak{g}^{\mu\nu}\frak{g}^{\lambda\alpha})
(2\frak{g}_{\beta\rho}\frak{g}_{\sigma\tau}-\frak{g}_{\rho\sigma}\frak{g}_{\beta\tau})
\frak{g}^{\beta\tau}_{\ \ ,\lambda}\frak{g}^{\rho\sigma}_{\ \ ,\alpha}\right]\,.
\end{eqnarray}
Here $\frak{t}_{LL}^{\mu\nu}=(-\frak{g})t_{LL}^{\mu\nu}$, where 
$t^{\mu\nu}_{LL}$ denotes the Landau-Lifshitz pseudo-tensor \cite{notation}.
For $\frak{g}\ne0$ we may substitute  Eqs. (\ref{invrelation}) 
into the equations of motion (\ref{efe}) and recover the standard Einstein's 
field equations depending on the metric. In this case,  
Eq. (\ref{efe}) implies that 
$\frak{T}^{\mu\nu}_{total}$ satisfies the Landau-Lifshitz energy-momentum conservation law
\be\label{conservation}
\partial_{\nu}\frak{T}^{\mu\nu}_{total} =0\,.
\ee

Let us consider a vacuum singularity, e.g. the Schwarzschild singularity, where $\frak{g}^{\mu\nu}$ is finite but degenerate. 
Notice that some of the terms in  
$\frak{t}^{\mu\nu}_{LL}$ have negative powers of $\frak{g}$. These terms become ambiguous  when the metric density 
becomes degenerate. This does not mean that
 $\frak{t}^{\mu\nu}_{LL}$ must diverge at singularity, e.g. in the Schwarzschild  coordinates 
$\frak{t}^{\mu\nu}_{LL}$ has a well defined limit as the singularity is approached.   
Still the ambiguity in $\frak{t}^{\mu\nu}_{LL}$ may give rise to 
difficulties in the extension of the solution across the singularity by using 
Eq. (\ref{efe}) alone.  
To overcome these difficulties  
we shall postulate new junction conditions at the singularity. 

Our goal is to find 
junction conditions that can be combined with the equations of motion (\ref{efe})
in a manner that would provide a unique extension of the solution 
across the hypersurface 
where the singularity resides. We would like to exploit the fact that 
the components of the metric density and their derivatives may  
remain finite
as the singularity is approached, and so it is possible to demand that 
these components would have some degree of smoothness at the singularity. 
It is tempting to demand the components of the metric density 
would be analytical functions of the coordinates. 
In fact requiring analyticity of the metric is a standard 
method to continue the Kerr geometry through its ring singularity (see e.g. \cite{he}).
However, in some respect demanding analyticity of the metric density is a too strong requirements. 
An analytical continuation 
completely determines the extension of the solution, so it 
is no longer necessary to solve the equations of motion. 
This however contradicts our goal
of keeping the standard physical picture 
where the solution is evolved via differential equations, and the junction conditions merely
determines how the solution is extended across the hypersurface with the singularity.

A weaker condition than analyticity is smoothness (i.e. $C^\infty$). Indeed an assumption of a smooth metric 
is often found in standard GR theorems.
For example, in a rigorous initial value formulation one assumes smooth initial data \cite{Wald}.   
In analogy with this standard GR assumption, we 
 postulate that the components of metric density 
  be smooth functions of the coordinates.
More precisely, we focus attention to metric density configurations that satisfy Eq. (\ref{efe}), not necessarily in vacuum, where $\frak{g}^{\alpha\beta}$ 
is smooth away from a genuine singularity of the geometry, and its components  and all of their partial derivatives have a well defined limit as the singularity is approached. 
We assume that $\frak{g}$ may be zero 
at most on a hypersurface, but not in an open set (i.e. not in the bulk), 
 and demand 
  that at a singular hypersurface, where $\frak{g}=0$, the components of  $\frak{g}^{\mu\nu}$ remain smooth. 
We then use these conditions together with Eq. (\ref{efe}) to extend the 
solution across the singularity. 
Let us now discuss few examples in detail.

\section {Examples}\label{examples}

\subsection {Friedmann-Robertson-Walker}
 First  we consider 
 a flat FRW Universe filled with a perfect fluid.
In the standard GR formulation this solution is nonextendible since its geometry becomes singular  
  at the big bang. 
By virue of the underlying symmetry we express 
the metric density as $\frak{g}^{\alpha\beta}=A(\eta)\eta^{\alpha\beta}$, where $\eta^{\mu\nu}$
denotes the Minkowski metric, and the scale factor is given by $a=\sqrt{A}$, where $A\ge 0$. As in GR we fix the signature
of the metric to be, say  $(-,+,+,+)$. 
 The matter source term reads 
$\frak{T}^{\mu\nu}={\rm diag}(\tilde{\rho},\tilde{p},\tilde{p},\tilde{p})$, 
where $p=A^{-3}\tilde{p}$ and $\rho=A^{-3}\tilde{\rho}$ are the proper pressure and enregy density, 
respectively.
For $A\ne0$ we 
substitute the expressions for $\frak{T}^{\mu\nu}$ and $\frak{g}^{\mu\nu}$ into Eq. (\ref{efe}) and obtain 
\begin{eqnarray}\label{forder}
&&3\dot{A}^2={32\pi}\tilde{\rho}\\\label{sorder}
&&3\dot{A}^2-4A\ddot{A}=32\pi \tilde{p}\,.
\end{eqnarray}
Here an overdot denotes differentiation with respect to conformal time $\eta$. 
We assume an equation of state of the form ${p}=w{\rho}$, where $w$ is a constant.
Substituting this relation into Eqs. (\ref{forder},\ref{sorder}) gives 
 $\tilde{\rho}\propto A^{3(1-w)/2}$. We substitute $\tilde{\rho}(A)$ into Eq. (\ref{forder}) and solve for $A$.
Recall that we seek a solution where $\frak{g}^{\alpha\beta}$ and all of  its partial derivatives 
have a well definied limit as singularity is approached. For the paricular 
coordinates in use, this condition is satsified for a radiation dominated Universe charterized by $w=1/3$.
Demanding that $\frak{g}^{\alpha\beta}$ be smooth at the singularity gives 
\be\label{frw}
A\propto \eta^2\,.
\ee
This solution describes a bounce at the big bang.

Notice that the matter source term $\frak{T}^{\mu\nu}$ is continuous at the singularity, but
the complete source term $\frak{T}_{total}^{\mu\nu}$ has an ambiguity at $A=0$, since 
$\frak{t}^{\mu\nu}_{LL}$ contains the combination 
$A/A$, which is ambiguous for $A=0$. 
Imposing continuity, we define $\frak{t}_{LL}^{\mu\nu}(0)$ to be the limit of $\frak{t}_{LL}^{\mu\nu}(\eta)$ as $\eta\rightarrow 0$. 
This gives, for all values of $\eta$, $16\pi\frak{t}_{LL}^{\mu\nu}=-\dot{A}^2{\rm diag}(3/2,7/2,7/2,7/2)$. 
Substituting this expression into the right hand side of Eq. (\ref{efe}) shows that solution (\ref{frw}) satisfies 
 the equation of motion (\ref{efe}) at the singularity. 
Moreover, evaluating the divergence of $\frak{T}_{total}^{\mu\nu}$ 
shows that the energy-momentum conservation 
law (\ref{conservation})  is satisfied {\em at the singularity}.
By contrast, it is hard to make sense of the covariant conservation
 law $\nabla_{\alpha}T^{\beta\alpha}=0$ at the big bang,  
since  both $\nabla_{\alpha}$ and $T^{\beta\gamma}$ diverge at the singularity. 

The above solution shows that the FRW singularity is described by a field configuration 
where both
 $\frak{g}^{\alpha\beta}$ and $\frak{g}_{\alpha\beta}$ are smooth, but the metric as defined in Eqs. (\ref{metricdef},\ref{ddmetricdef}) can not be constructed. 
This suggests the interpretation that a degenerate metric density 
 describes a pre-metric  configuration which is more primitive 
than a geometry.

\subsection {Black holes}

Next we study a spherically symmetric charged black hole characterized by a mass $M$ and a charge $Q$.
Substituting  the RN solution into Eq. (\ref{invrelation}) we find that for $r>0$ the non-zero components of the metric density are given by 
\begin{eqnarray}\label{RN}
&&\frak{g}^{tt}_{RN}=\frac{-r^4 \sin\theta}{r(r-2M)+Q^2}\ , \  \frak{g}^{\theta\theta}_{RN}=\sin \theta\ \\\nonumber
&& \frak{g}^{rr}_{RN}=[Q^2+r(r-2M)] \sin \theta \ , \ \frak{g}^{\phi\phi}_{RN}=(\sin \theta)^{-1}\,,
\end{eqnarray}
and $\sqrt{-\frak{g}_{RN}}=r^2\sin\theta$. 
We focus on the domain  $r<r_-$, where  $r_{\pm}=M\pm\sqrt{M^2-Q^2}$, and $M>|Q|$.
Notice that in the limit $r\rightarrow0$  the RN metric density 
becomes a non-divergent degenerate matrix. The $r=0$  surface  
corresponds to a genuine singularity since the scalar $R_{\mu\nu}R^{\mu\nu}$ diverges in the limit. 
We now extend the solution across this singularity to the domain $r<0$.
For this purpose, we first have to solve Maxwell equations in vacuum.
To overcome the singularity in the electromagnetic field we 
cast Maxwell equations in a densitized format, and obtain 
\be\label{maxwell}
\frak{F}^{\alpha\beta}_{\ \ ,\beta}=0\,.
\ee
Here $\frak{F}^{\alpha\beta}=\sqrt{-\frak{g}}F^{\alpha\beta}$, where $F^{\alpha\beta}$ 
is the electromagnetic field tensor. 
For $r>0$ the non zero components of the RN solution $\frak{F}^{\alpha\beta}_{RN}$ read
\be\label{maxsol}
\frak{F}_{RN}^{tr}=-\frak{F}_{RN}^{rt}=Q\sin \theta\,.
\ee
Notice that Eq. (\ref{maxwell}) has no singularity at $r=0$, and
solution (\ref{maxsol}) satisfies the Maxwell equations (\ref{maxwell}) in the entire domain  $-\infty<r<r_-$.
We now solve Eq. (\ref{efe}) in $r<0$.
Exploiting the underlying symmetry 
 we substitute $\frak{g}^{\alpha\beta}
={\rm diag}[-C^4(r)\sin\theta /D(r),D(r)\sin\theta,\sin\theta,(\sin\theta)^{-1}]$
into Eq. (\ref{efe}), where we introduced the combination $C^4/D$ in  $\frak{g}^{tt}$ 
to slightly simplify the equations.
Demanding that 
$\frak{g}^{rr},\frak{g}^{tt}$ be smooth at $r=0$, and using Eqs. (\ref{RN}) determines
 $C(r)$ and $D(r)$ uniquely. 
The complete solution for $-\infty<r<r_-$ is  given by Eq. (\ref{RN}).  This  expression for the 
extended geometry is also obtained for the
special case of the Schwarzschild solution 
where $Q=0$.
Figure 1 shows the Penrose-Carter diagram of the RN extension for $Q\ne 0$. We should mention that this 
extension had been previously conjectured in Ref. \cite{GKOS}.
Notice that the components of this 
extended RN metric density  
are analytical in the domain of interest, and in particular they are well defined at the $r=0$ singularity. However, 
the metric 
as defined in Eqs.   (\ref{metricdef},\ref{ddmetricdef}) can not be constructed at the singularity.
 Similar to the FRW example, we define $\frak{T}_{total}^{\mu\nu}(0)$ to be the limit of $\frak{T}_{total}^{\mu\nu}(r)$ as $r\rightarrow 0$.
With this definition, the equation of motion (\ref{efe}) and the conservation law (\ref{conservation}) 
are satisfied at the singularity.

\begin{figure}
 \includegraphics {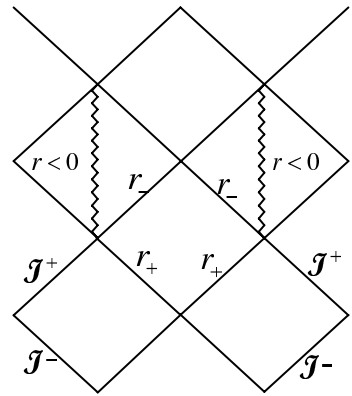}
  \caption{Penrose-Carter diagram of the extended RN spacetime.}
\label{fig1}
\end{figure}

The geometry of a rotating Kerr (and Kerr-Newmann) black hole can be analytically  
extended through its ring singularity 
using the standard GR formalism (see e.g. \cite{he}), and therefore a regularization of the singularity 
is not essential in this case.   
Nevertheless, it interesting to note that in the Boyer-Lindquist coordinates 
 the components of the metric density of the Kerr (and Kerr-Newmann) solution are analytical 
in the neighborhood of the singularity. Here again the metric density takes the form of a 
degenerate non-diverging matrix at the singularity. 

\subsection{Kasner}

The Kasner solution is a vacuum solution which is 
flat and homogenous but anisotropic. In the Kasner coordinates $(t,x,y,z)$ 
the metric density of the Kasner solution is given by 
\begin{equation}\label{kasner}
{\frak g}^{\alpha\beta}_{Kas}={\rm diag}(-t,t^{1-2p_1},t^{1-2p_2},t^{1-2p_3})\,.
\end{equation}
Here  the three parameters $p_i$, satisfy
\begin{equation}\label{constraints}
p_1+p_2+p_3=p_1^2+p_2^2+p_3^2=1\,.
\end{equation}
The Kasner geometry is regular for $t>0$ but has a singularity at $t=0$. 
To be able to use our model and extend the geometry to the domain $t<0$, 
we should check that the components of 
the metric density, and their partial derivatives, have a well defined limit as the singularity is approached.
Here however there is a difficulty. 
Notice that the constraints (\ref{constraints}) imply that at least one of the $p_i$ parameters  is greater than $1/2$, and
so at least one of the components of ${\frak g}^{\alpha\beta}_{Kas}$ must diverge at the singularity.
To resolve this difficulty we should seek a coordinate transformation that removes the singularity from  ${\frak g}^{\alpha\beta}_{Kas}$.
Below we provide such a transformation for the case where the $p_i$  parameters are given by 
three different rational numbers.

Let us consider first the domain $t>0$ and introduce 
 the coordinate transformation $t=\epsilon\eta^s$, where $s$ is a positive integer whose value is specified below, and $\epsilon=1$. 
 In the new coordinates $(\eta,x,y,z)$ the metric density reads
\begin{equation}\label{kasner2}
{\frak g}^{\alpha'\beta'}_{Kas}={\rm diag}(-s^{-1}\eta,s\eta^{q_1},s\eta^{q_2},s\eta^{q_3})\,.
\end{equation}
Here $q_i=2s(1-p_i)-1$. Let us denote the combinations $(1-p_i)$ with $m_i/n_i$, where $m_i,n_i$ are
positive integers.  
Setting $s=n_1n_2n_3$ implies that the parameters $q_1,q_2,q_3$ are given by positive odd integers.
Notice that the components of the Kasner metric density ${\frak g}^{\alpha'\beta'}_{Kas}$ and their partial derivatives, both have a well 
defined limit as $\eta\rightarrow 0$ as desired.

We are now ready to use our model and extend the solution to  $\eta<0$. 
Notice that solution (\ref{kasner2}) satisfies Eq. (\ref{efe}) in $\eta<0$, 
though here the signature is $(+,-,-,-)$. Imposing our junction conditions by demanding that the components of the metric density be smooth at the $\eta=0$ singularity, ensures that the parameters  
$s$ and $q_1,q_2,q_3$ have the same values for $\eta<0$ and $\eta>0$. This uniquely determines the continuation of the geometry across the singularity.  
It is now possible to transform back to the original coordinates (For the case where $s$ is an even number, we set $\epsilon=-1$ for $\eta<0$, so that negative values of $\eta$ correspond to negative values of $t$).

\section {Conclusions}\label{conclusions}

Singularities mark the breakdown of the laws of general relativity.  
There is a widely accepted view that a viable quantum theory of gravity should have a mechanism that removes singularities, for example by some quantum-gravity 
phenomena that smooth them out. In this paper we have presented a classical model that illustrates an alternative point of view. We showed that it is possible to keep the singularities in the solution, so that they become a legitimate part of the physical model.  

The construction of our model 
 was based on the observation that the metric density encodes the same amount of information as the metric,  
but unlike the metric, its components and their partial derivatives may remain finite at
a class of spacetime singularities. Our strategy was to 
exploit this property, and replace the metric with the metric density as the fundamental variable of gravity. By retaining the dynamical evolution of GR, and supplementing it with junction conditions we showed
that it is possible to extend certain singular geometries across their singularities.

It would be interesting to see if our model 
could be used to extend singular geometries in $d$ dimensions. 
Curiously, however, not all the observables  
can be adjusted to accommodate a $d$ dimensional metric density. 
In particular notice that $\frak{g}_{\alpha\beta}$ transforms as a tensor density of weight $-{(d-3)}$, and so
 the inner product 
$(u^\alpha,v^\beta)$  defined by Eq. (\ref{uu})
is a scalar only  if $d=4$.

\acknowledgments 

I thank E. E. Flanagan, I. M. Wasserman, A. Ori, and S. Teukolsky   for discussions. 
This work was supported by NSF grants, PHY-0757735, PHY-0555216.

\newcommand{\apjl}{Astrophys. J. Lett.}
\newcommand{\aap}{Astron. and Astrophys.}
\newcommand{\cmp}{Commun. Math. Phys.}
\newcommand{\grg}{Gen. Rel. Grav.}
\newcommand{\lr}{Living Reviews in Relativity}
\newcommand{\mnras}{Mon. Not. Roy. Astr. Soc.}
\newcommand{\pr}{Phys. Rev.}
\newcommand{\prsl}{Proc. R. Soc. Lond. A}
\newcommand{\ptrsl}{Phil. Trans. Roy. Soc. London}


\begin{thebibliography}{10}
\bibitem{adjoint}
This property follows from the 
transformation law for  $\frak{g}^{\alpha\beta}$ together with the following properties of the adjoint 
$adj(AB)=adj(A)adj(B)$, where $A,B$  are matrices, 
and  $adj(C)=C^{-1}\det(C)$ where $C$ 
is an invertable matrix (notice that the coordinate transformation matrix $\frac{\partial {x'}^{\alpha}} {\partial x^{\beta}}$
is invertable).
\end{thebibliography}
\end{document}